\documentclass[runningheads]{llncs}
\makeatletter
\renewcommand*{\@fnsymbol}[1]{\ifcase#1\or* \else*\fi}
\makeatother
\usepackage[T1]{fontenc}
\usepackage{graphicx}
\usepackage{subfig}
\usepackage{amsmath}
\usepackage{amsfonts}
\usepackage{url}
\usepackage{multirow}
\usepackage{paralist}
\usepackage{algorithm}
\usepackage{algpseudocode}
\usepackage{color}
\usepackage{bm}
\usepackage{comment}
\newtheorem{mydef}{Definition}
\newtheorem{myprob}{Problem}
\usepackage{amssymb}
\usepackage{booktabs, makecell, tabularx}
\usepackage[export]{adjustbox}
\usepackage{mathtools}
\usepackage{wrapfig}
\usepackage{footnote} 
\usepackage{rotating}

\makeatletter
\newcommand{\printfnsymbol}[1]{%
  \textsuperscript{\@fnsymbol{#1}}%
}
\makeatother
\providecommand{\abs}[1]{\lvert#1\rvert} 
\usepackage[misc]{ifsym}

\begin{document}
\title{Hop-Count Based Self-Supervised Anomaly
Detection on Attributed Networks}
\titlerunning{HCM based Self-Supervised Anomaly
Detection on Attributed Networks}
%

\author{Tianjin Huang $^{*}$ \Letter  \and
Yulong Pei \thanks{Both authors contributed equally to this research.}\and
Vlado Menkovski \and
Mykola Pechenizkiy}

\authorrunning{T. Huang et al.}
%
\institute{Department of Mathematics and Computer Science, Eindhoven University of Technology, 5600 MB Eindhoven, the Netherlands\\
\email{\{t.huang,y.pei.1,v.menkovski, m.pechenizkiy\}@tue.nl}}
\maketitle              
\begin{abstract}
A number of approaches for anomaly detection on attributed networks have been proposed. However, most of them suffer from two major limitations: (1) they rely on unsupervised approaches which are intrinsically less effective due to the lack of supervisory signals of what information is relevant for capturing anomalies, and (2) they rely only on using local, e.g., one- or two-hop away node neighbourhood information, but ignore the more global context. Since anomalous nodes differ from normal nodes in structures and attributes, it is intuitive that the distance between anomalous nodes and their neighbors should be larger than that between normal nodes and their (also normal) neighbors if we remove the edges connecting anomalous and normal nodes. Thus, estimating hop counts based on both global and local contextual information can help us to construct an anomaly indicator. Following this intuition, we propose a hop-count based model (HCM) that achieves that. Our approach includes two important learning components: (1)~Self-supervised learning task of predicting the shortest path length between a pair of nodes, and (2)~Bayesian learning to train HCM for capturing uncertainty in learned parameters and avoiding overfitting. Extensive experiments on real-world attributed networks demonstrate that HCM consistently outperforms state-of-the-art approaches.
\end{abstract}
\section{Introduction}
Attributed networks are ubiquitous in a variety of real-world applications. Attributed networks can be utilized to represent data from different domains. For example, in a social network, each node can represent a user, an edge denotes the friend relation between users, and user profiles are the attributes to describe users. A citation network consists of papers as the nodes, citation relations as the edges, and words in paper abstracts can be the attributes of papers. Unlike plain networks where only structural information exists, attributed networks also contain rich features to provide more details to describe (elements of) networks. Due to the ubiquity of attributed networks, various tasks on attributed networks have been widely studied such as community detection~\cite{falih2018community,pei2015nonnegative}, link prediction~\cite{brochier2019link,li2018streaming} and network embedding~\cite{huang2017accelerated,meng2019co}.

Among these tasks on attributed networks, anomaly detection is perhaps one of the most important ones in the current analytics tasks -- it can shed light on a wide range of real-world applications such as fraud detection in finance and spammers discovery in social media. 

\begin{wrapfigure}{r}{0.5\linewidth}
\vspace{-0.4in} 
	\begin{center}
		\includegraphics[width=0.5\textwidth]{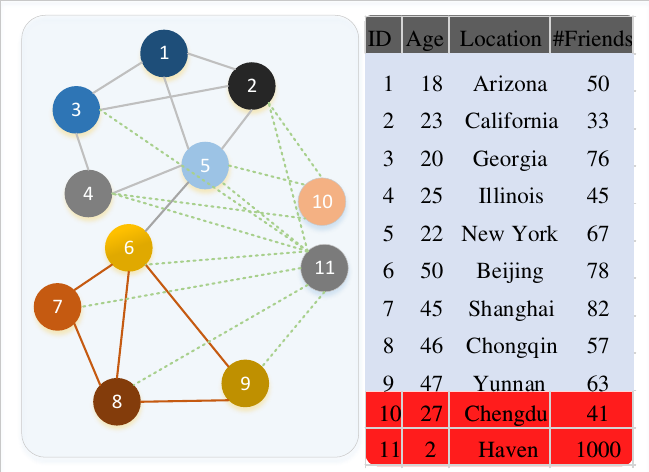}
		\caption{A toy example for illustrating anomalies on attributed networks.}
		\label{fig:example}
	\end{center}
\vspace{-0.3in}
\end{wrapfigure}

Unlike in anomaly detection on plain networks, detecting anomalies on attributed networks should rely on two sources of information: (1)~the structural patterns of how nodes interconnect or interact with each other, which are reflected by the topological structures, (2)~the distributions of nodal features. Therefore, it is more challenging to detect anomalies on attributed networks. Fig.~\ref{fig:example} shows a toy example of anomalies on an attributed network. In Fig.~\ref{fig:example}, nodes represent the individuals and links represent the connections. Node 10 can be identified as an attribute anomaly since it connects to the community with US cities while its location attribute is a China City. Node 11 can be identified as a structural anomaly since it connects to almost all other nodes where some connections are unreasonable.  

To solve this challenging problem, various supervised and unsupervised approaches have been proposed recently. Unsupervised anomaly detection methods are preferable in practice because of the prohibitive cost for accessing the ground-truth anomalies~\cite{ding2019deep}. Hence, in this study, we focus on unsupervised fashion to detect anomalies in the sense that we do not require access to ground truth anomalies in the training data. Intrinsically, unsupervised learning approaches demonstrate worse performance than supervised learning approaches (when they are applicable) due to the lack of supervisory signals. Recently, various self-supervised techniques have been proposed to learn a better feature representation via designing supervisory signals directly from input data. It has been shown, e.g.,\ in the computer vision domain, that self-supervised learning can greatly benefit model's robustness and out-of-distribution detection~\cite{hendrycks2019using}.

In this paper, we employ the idea of self-supervised learning by forcing the prediction of the shortest path length between pairs of nodes (denoted as hop counts for convenience) to detect anomalies on attributed networks. This idea allows to address another limitation of existing approaches that utilize only the local context of nodes to detect anomalous nodes while neglecting the global context information that provides a complementary view of the network structure patterns including anomalous structures. It has been shown that the global context of a node is helpful to learn node representation that can finely characterize the similarity and differentiation between nodes~\cite{peng2020self}. 
Therefore, it is reasonable to assume that effectively anomaly detection methods should consider local and global contextual information.

\subsection{Our HCM approach in a nutshell.}
We use hop count prediction as a self-supervised learning task to capture local and global contextual information for node representation. We use GCN to learn node representations for refining the graph. Then the learned node representation is used to predict hop counts for arbitrary node pairs. We design two anomaly scores based on the intuition that the true hop counts between anomalous nodes and their neighbors shall be larger than that between normal nodes and their neighbors, e.g.\ in our toy example in Fig.~\ref{fig:example}, the hop counts between node 11 and their neighbors should be larger than 1 since node 11 is probably from the community with reddish nodes.

We adopt Bayesian learning to train our HCM model because Bayesian methods are appealing in their ability to capture uncertainty in learned parameters and avoid overfitting~\cite{welling2011bayesian}. Specifically, we exploit Stochastic Gradient Langevin Dynamics (SGLD)~\cite{welling2011bayesian} to optimize our model and conduct Bayesian inference. 

\subsection{Summary of the contributions}
The contributions of this paper are summarized as follows:
\begin{itemize}
    \item We employ self-supervised learning ideas in the graph anomaly detection domain and make use of both global and local contextual information of nodes, i.e., hop counts, to detect anomalies on attributed networks. 
    \item With the help of the self-supervised learning technique, we propose our HCM model to learn node representations capturing local and global contextual information of nodes. And based on the HCM model, we design two anomaly scores to detect anomalies. Experimental results demonstrate the effectiveness of our approach.
    \item We exploit SGLD to optimize HCM model for capturing uncertainty in learned parameters and avoid overfitting. We experimentally demonstrate that our model optimized by SGLD performs better than SGD in anomaly detection. Besides, SGLD achieves a steadier behaviour during the training process.   
\end{itemize}
\section{Related Work}
Anomaly detection is an important task in data mining and machine learning. Previous studies roughly can be categorized into four types~\cite{ding2019deep}: community analysis, subspace selection, residual analysis and deep learning methods. Community analysis methods~\cite{breunig2000lof,gao2010community,gutierrez2019multi} detect anomalies by identifying current node's abnormality with other nodes within the same community. Subspace selection approaches~\cite{perozzi2016scalable,perozzi2014focused} learn a subspace for features and then discover anomalies in the learned subspace. Residual analysis methods~\cite{li2017radar,peng2018anomalous} explicitly model the residual information by reconstructing the input attributed network based on matrix factorization. With the popularity of deep learning techniques, methods using deep neural networks such as graph convolutional networks (GCN) and network embedding to  detect anomalies have been proposed~\cite{ding2019deep,liang2018semi,pei2020resgcn}. Among these methods, seven popular methods are chosen as baselines in this paper including LOF~\cite{breunig2000lof}, AMEN~\cite{perozzi2016scalable}, Radar~\cite{li2017radar}, ANOMALOUS~\cite{peng2018anomalous},  DOMINANT~\cite{ding2019deep}, MADAN~\cite{gutierrez2019multi} and ResGCN~\cite{pei2020resgcn}. LOF~\cite{breunig2000lof} measures how isolated the object is with respect to the surrounding neighborhood and detects anomalies at the contextual level. LOF only considers nodal attributes. AMEN~\cite{perozzi2016scalable} analyzes the abnormality of each node from the ego-network point of view. Radar~\cite{li2017radar} detects anomalies by characterizing the residuals of attribute information and its coherence with network information. ANOMALOUS~\cite{peng2018anomalous} is a joint anomaly detection framework to select attributes and detect anomalies using CUR decomposition of a matrix. DOMINANT~\cite{ding2019deep} selects anomalies by ranking the reconstruction errors where the errors are learned by GCN. MADAN~\cite{gutierrez2019multi} uses the heat kernel as filtering operator to exploit the link with the Markov stability to find the context for multi-scale anomalous nodes. ResGCN~\cite{pei2020resgcn} learns the residual information using a deep neural network, and reduces the adverse effect from anomalous nodes using the residual-based attention mechanism. 

\section{Problem Definition}
\label{prob}
Following the commonly used notations, we use bold uppercase characters for matrices, e.g., $\bm{X}$, bold lowercase characters for vectors, e.g., $\bm{b}$, and normal lowercase characters for scalars, e.g., $c$. The $i^{th}$ row of a matrix $\bm{X}$ is denoted by $\bm{X}_{i,:}$ and $(i,j)^{th}$ element of matrix $\bm{X}$ is denoted as $\bm{X}_{i,j}$. The Frobenius and $L_2$ norm of a matrix are represented as $\|\cdot\|_F$ and $\|\cdot\|_2$ respectively. The number of elements of a set is denoted by $\abs{\cdot}$. 

\begin{mydef}
\textbf{Attributed Networks}. An attributed network $\mathcal{G} = \{V,E,\bm{X}\}$ consists of: (1) a set of nodes $V=\{v_1,v_2,...,v_n\}$, where $|V|=n$ is the number of nodes; (2) a set of edges $E$, where $|E|=m$ is the number of edges; and (3) the node attribute matrix $\bm{X}\in \mathbb{R}^{n\times d}$, the $i^{th}$ row vector $\bm{X}_{i,:}\in\mathbb{R}^{d}$ is the attribute of node $v_i$ where $d$ is the number of attributes.
\end{mydef}

The topological structure of attributed network $\mathcal{G}$ can be represented by an adjacency matrix $\bm{A}$, where $\bm{A}_{i,j}=1$ if there is an edge between node $v_i$ and node $v_j$. Otherwise, $\bm{A}_{i,j}=0$. We focus on the undirected networks in this study and it is trivial to extend it to directed networks. The attribute of $\mathcal{G}$ can be represented by an attribute matrix $\bm{X}$. Thus, the attributed network can be represented as $\mathcal{G} = \{\bm{A},\bm{X}\}$. With these notations and definitions, same to previous studies~\cite{ding2019deep,li2017radar,pei2020resgcn,peng2018anomalous}, we formulate the task of anomaly detection on attributed networks:
\begin{myprob}
\textbf{Anomaly Detection on Attributed Networks}. Given an attributed network $\mathcal{G} = \{\bm{A},\bm{X}\}$, which is represented by the adjacency matrix $\bm{A}$ and attribute matrix $\bm{X}$, the task of anomaly detection is to find a set of nodes that are rare and differ singularly from the majority reference nodes of the input network.
\end{myprob}

\section{Hop-Count based Model}
\label{model}
In this section we first present a detailed description for our proposed HCM model. Then the SGLD optimization strategy will be introduced. Finally, we discuss the anomaly scores based on different strategies.

\begin{figure*}
	\begin{center}
		\includegraphics[width=0.9\textwidth]{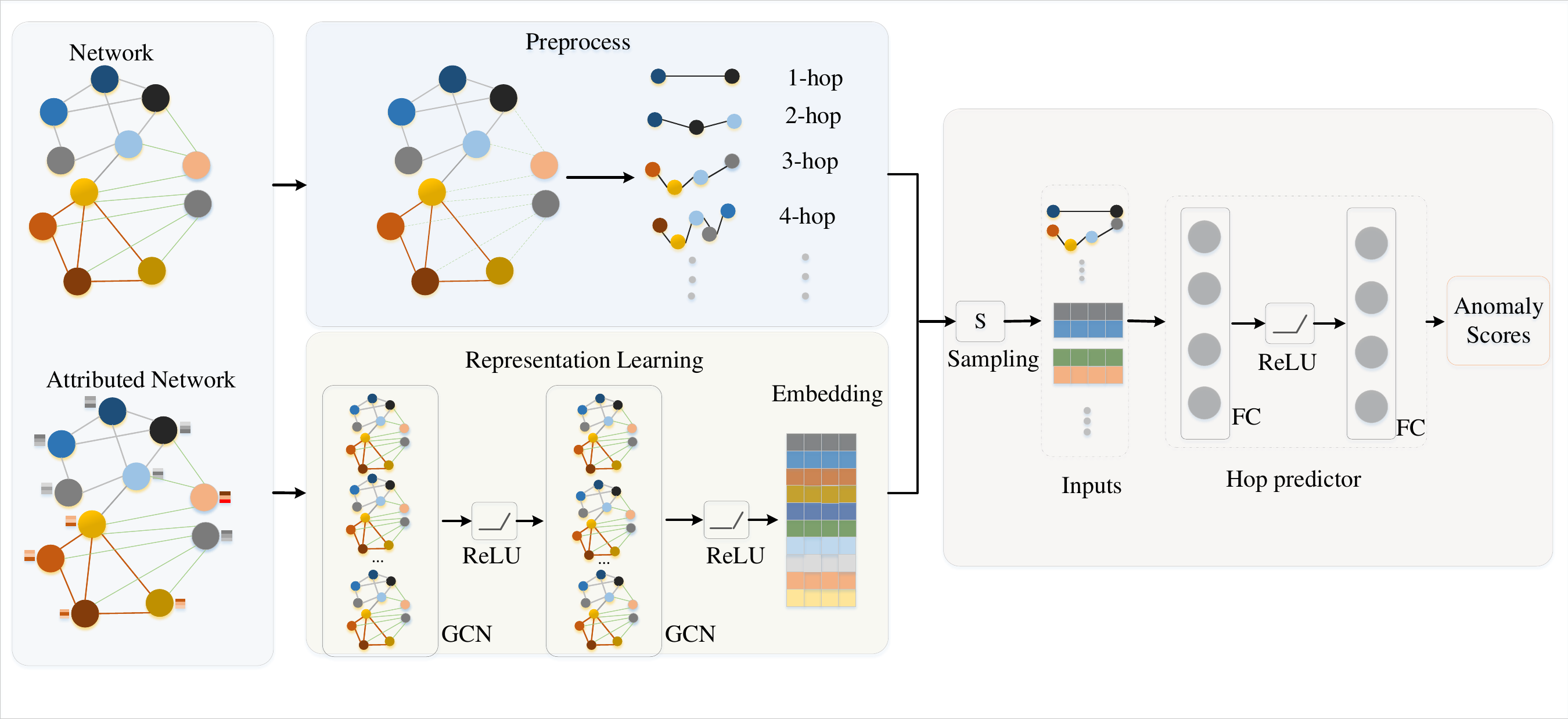}
		\caption{The framework of our proposed HCM model. The dotted lines in Preprocess component denote the dropped edges.}
		\label{fig:frame}
	\end{center}
\vspace{-0.5in}
\end{figure*}

\subsection{Model Framework}

In this section, we introduce the framework of our proposed HCM model in detail. HCM model is based on self-supervised task designed by some innate characteristics of the data. Recently, several self-supervised tasks have been proposed for graph data. For example, node clustering task where a clustering operation will be used to provide pseudo labels; Graph partition task where graph partitioning will provide pseudo labels~\cite{you2020does}; PairDistance task where hop counts of node pairs are used as label~\cite{jin2020self}; PairwiseAttrSim where the similarities of node pairs are taken as label~\cite{jin2020self}. Here we adopt PairDistance as our self-supervised task. There are three reasons accounting for using PairDistance as our self-supervised task: 1) It is cheap to get the true label during the training process; 2) The hop counts capture both global and local information of a network; 3) hop counts can be utilized to directly construct anomaly scores of nodes. Our HCM model is composed of three components (Fig.~\ref{fig:frame}): Preprocess components, Graph Convolutional network (GCN) and Multilayer perception (MLP). Preprocess is to generate labels during the training process; GCN is to learn the representation of the attributes networks; MLP is the classifier to predict hop counts for node pairs.  
\subsubsection{Preprocess} There are two operations in Preprocess component. The first is to drop edges with small similarities in the adjacency matrix. The second is to generate labels, namely the hop counts of all node pairs.

\textbf{Drop Edges} Considering that the anomalies on the attribute network could potentially perturb links, we remove some edges in the adjacency matrix before utilizing it to generate labels. Specifically, we remove the potential perturbed links according to the similarities of node pairs. \cite{wu2019adversarial} has demonstrated that removing links with small similarities can effectively purify perturbed links, especially for adversarial perturbed links. Similar to \cite{wu2019adversarial}, for binary features, we adopt the Jaccard similarity to measure the similarities:
\begin{align}
    S_{Jar}(i,j)=\frac{\abs{\bm{X}_{i,:} \cap \bm{X}_{j,:}}}{\abs{\bm{X}_{i,:} \cup \bm{X}_{j,:}}}.
    \label{jar}
\end{align}
For continuous features, we adopt the Cosine similarity to measure the similarities:
\begin{align}
    S_{Sin}(i,j)=\frac{\bm{X}_{i,:}^T \bm{X}_{j,:}}{\lVert \bm{X}_{i,:}\rVert_2 \lVert \bm{X}_{j,:}\rVert_2}.
    \label{Sim}
\end{align}
We use the hyper-parameter drop ratio $R$ to control  the percentage of links that will be removed.

\textbf{Generate Labels} After dropping some edges, the modified adjacency matrix will be used to generate labels. Specifically, we firstly utilize Dijkstra search algorithm to calculate the hop counts for all node pairs: $Y_{hop}^1=\{(v_m,v_n)\lvert hop(v_m,v_n)$ $= 1, v_m,v_n \in V \}$, $Y_{hop}^2=\{ (v_m,v_n) \lvert hop(v_m,v_n)=2,v_m,v_n \in V \}$, $Y_{hop}^3= \{(v_m,v_n)$ $\lvert hop(v_m,v_n)=3,v_m,v_n \in V \}$..., $Y_{hop}^i=\{ (v_m,v_n)$ $\lvert hop(v_m,v_n)=i,v_m,v_n \in V \}$, where $hop(v_m,v_n)$ denotes the hop counts between node $v_m$ and $v_n$. Then these hop counts will be grouped into $C$ classes: $hop(v_m,v_n)=1,hop(v_m,v_n)=2,...,hop(v_m,v_n)>=C$. Besides, Considering that the $\lvert Y_{hop}^i \rvert$ are different for each set $Y_{hop}^i$, which will lead to imbalanced training set. In order to avoid the bias induced by imbalanced training set, we sample a same amount of node pairs from each set $Y_{hop}^i$  during the training, the amount of node pairs is decided by the smallest $\lvert Y_{hop}^i \rvert \cdot S$ where $S$ is sampling ratio. 

\subsubsection{GCN}
GCN learns node representations by passing and aggregating messages between neighboring nodes. Different types of GCN have been proposed recently~\cite{kipf2016semi,hamilton2017inductive}, and we focus on one of the most widely used versions proposed in~\cite{kipf2016semi}. Formally, a GCN is defined as
\begin{equation}
\label{gcn}
    \bm{h}^{(l+1)}_i=f\Big(\sum_{j\in Ne(i)}\frac{1}{\sqrt{\widetilde{\bm{D}}_{i,i}\widetilde{\bm{D}}_{j,j}}}\bm{h}^{(l)}_j \bm{W}^{(l)}\Big),
\end{equation}
where $\bm{h}^{(l)}_i$ is the latent representation of node $v_i$ in layer $l$, $Ne(i)$ is the set of neighbors of node $v_i$, and $\bm{W}^{l}$ is the layer-specific trainable weight matrix. $f(\cdot)$ is a non-linear activation function and we select ReLU as the activation function following previous studies~\cite{kipf2016semi} (written as $f_{ReLU}(\cdot)$ below). $\widetilde{\bm{D}}$ is the diagonal degree matrix of $\widetilde{\bm{A}}$ defined as $\widetilde{D}_{i,i}=\sum_{j}\widetilde{A}_{i,j}$ where $\widetilde{\bm{A}}=\bm{A}+\bm{I}$ is the adjacency matrix of the input attributed network $\bm{G}$ with self connections $\bm{I}$. Equivalently, we can rewrite GCN in a matrix form:
\begin{equation}
\label{gcn-mat}
	\bm{H}^{(l+1)}=f_{ReLU}\Big(\widetilde{\bm{D}}^{-\frac{1}{2}}\widetilde{\bm{A}}\widetilde{\bm{D}}^{-\frac{1}{2}}\bm{H}^{(l)}\bm{W}^{(l)}\Big).
\end{equation}
For the first layer, $\bm{H}^{(0)}=\bm{X}$ is the attribute matrix of the input network. Therefore, we have
\begin{equation}
\label{gcn-1st}
	\bm{H}^{(1)}=f_{ReLU}\Big(\widetilde{\bm{A}}\bm{X}\bm{W}^{(0)}\Big).
\end{equation}
In this paper, we denote the representation learned by the GCN as $\bm{Z}$.
\subsubsection{MLP}
The target of MLP component is to construct hop counts prediction task. MLP is composed of multiple fully connected layers. During training phase, CrossEntropy loss is used to construct supervisory signals. During inference, it predicts hop counts for any arbitrary node pair representations.  Given any node pairs $\bm{Z}_{m,:}$ and $\bm{Z}_{n,:}$, it can be formally expressed as following:
\begin{align}
    \bm{z}_{d}=\abs{\bm{Z}_{m,:}-\bm{Z}_{n,:}}\\
    F_{mlp}(v_{m},v_{n})=f_{ReLU}^{k}(\bm{z}_{d} \cdot \bm{W}),
\end{align}

where $F_{mlp}(v_{m},v_{n})$ is the hop counts prediction of node $v_m$ and $v_n$,  $f_{ReLU}^{k}$ is $k$ fully connected layers, $\bm{W}$ is total trainable weight matrix of MLP and $\abs{\cdot}$ here is the absolute function. The reason for using absolution function is based on the consideration that the hop counts from node $\bm{Z}_{m,:}$ to $\bm{Z}_{n,:}$ and from node $\bm{Z}_{n,:}$ to $\bm{Z}_{m,:}$ are the same on undirected networks. For directed networks, absolution function can be removed.

For the convenience of formulation, we denote the whole model as follows:
\begin{align}
        F_{\bm{w}}(\bm{X},\bm{A})=F_{mlp}(v_{m},v_{n}) \circ H^l,
\end{align}
where $\bm{w} \in \Omega$ are the parameters of the HCM model and it contains parameters from GCN and MLP components.

\subsubsection{The Self-supervised Learning Loss}
As described above, we adopt PairDistance as the self-supervised task. PairDistance task is to predict hop counts for any given node pairs. And it can be formulated as a multi-class classification problem. Therefore, we formulate the self-supervised learning loss as follows:
\begin{align}
\mathcal{L}_{self}(\bm{w},\bm{A},\bm{X})=\frac{1}{\lvert \mathcal{S} \rvert} \sum_{(v_m,v_j) \in \mathcal{S}} L(F_{mlp}(v_{m},v_{n}), hop(v_m,v_n)),
\end{align}
where $\mathcal{S}$ is a set of randomly sampled node pairs and $L(\cdot)$ is the CrossEntropy loss.

\subsection{Model Training}

The Bayesian framework suggests that the parameters of a model are random variables instead of determined. Therefore, it is crucial to estimate the distribution of model parameters. According to Bayes' theorem, the posterior distribution of the parameters can be defined as follows:
\begin{align}
    p(\bm{w}\lvert \bm{X},\bm{A},\bm{Y_{hop})}=\frac{p(\bm{Y_{hop}}\lvert  \bm{X},\bm{A},\bm{w})p(\bm{w})}{p(\bm{Y_{hop}}\lvert \bm{X},\bm{A})},
    \label{bayesian}
\end{align}
where $p(\bm{w})$ is a prior distribution for the model's parameters $\bm{w}$. 

According to Eq.\eqref{bayesian}, there are two different approaches to do the inference:(1) Maximum posterior probability (MAP) estimation finds the mode of the posterior distribution. (2) Bayesian inference computes the posterior distribution itself. Considering that the MAP-estimation can not capture the model's uncertainty, we choose the second approach as our solution. The Bayesian inference of outputs is given as followings:
\begin{align}
    \small
    p(y_{hop}^* \lvert x^*,\bm{X},\bm{A},\bm{Y_{hop}})=\int_\Omega{p(y^*\lvert x^*,\bm{w})p(\bm{w}\lvert \bm{X},\bm{A},\bm{Y_{hop}})d\bm{w}},
    \label{bayesian inference}
\end{align}
Due to the integration, it is impossible to achieve the prediction by computing the Eq.\eqref{bayesian inference} directly. Fortunately, Stochastic gradient Langevin dynamics (SGLD)\cite{welling2011bayesian}  provides a general framework to solve Eq. \eqref{bayesian inference}. The SGLD update is defined as follows:
\begin{align}
    \delta_{\bm{w}_{t}}=\frac{\epsilon_{t}}{2}(\nabla_{\bm{w}_{t}}log p(y_{hop}\lvert \bm{w}_{t})+\nabla_{\bm{w}}log p(\bm{w}_{t}))+\eta_t \\
    \eta_t \in N(0,\epsilon)\\
    \bm{w}_{t+1}=\bm{w}_{t}-\delta_{\bm{w}_{t}},
\end{align}
where $\epsilon_{t}$ is the step size. The log prior term will be implemented as weight decay.~\cite{welling2011bayesian} has shown that under suitable conditions, e.g., $\sum \epsilon_t=\infty$ and $\sum \epsilon_t^2 \textless \infty$ and others, $\bm{w}$ converges to the posterior distribution. The pseudocode of HCM model training are summarized in Algorithm~\ref{alg:1} (In Appendix~\ref{algo}).

\subsection{Model Inference for Anomaly Detection}
\subsubsection{Inference}
Based on SGLD update, the posterior inference can be achieved by adding Gaussian noise to the gradients at each step  and the prediction can be estimated by the posterior sample averages after a "burn in" phase~\cite{welling2011bayesian}:
\begin{align}
    \widetilde{Y}_{hop}(v_i,v_j)=\frac{1}{T}\sum_{t=1}^{T} F_{\bm{w}_{t}}(v_i,v_j \lvert \bm{X},\bm{A})
\end{align}

\subsubsection{Anomaly Scores}
Based on the HCM model, we design two anomaly criteria for catching anomalies. The first anomaly criterion is the averaged predicted hop counts between a node and its neighbors. Since anomalous nodes are different from their local neighbors, the predicted hop counts are expected to be larger than normal nodes. Therefore, the higher averaged predicted hop counts, the higher probability of being anomalous nodes. The second anomaly criterion combines the averaged predicted hop counts and the inferred variance of posterior samples.~\cite{kendall2017uncertainties} shows the evidence that the lack of data could increase the model's uncertainty. Since anomalies are rare on the network, the inferred variance on anomalous nodes tends to be higher. 

\textbf{Average Hop Prediction (AHP)} which is the averaged hop predictions between a node and its neighbors. The expression is defined as followings:
\begin{align}
    S_{AHP}(v_i)=\frac{\sum_{v_j} \widetilde{Y}_{hop}(v_i,v_j)}{\lvert \mathcal{N}(v_i) \rvert}, v_j \in \mathcal{N}(v_i),
    \label{ahp_i}
\end{align}
where $\mathcal{N}(v_i)$ is the neighbors of node $v_i$. $S_{AHP}(v_i)$ represents the averaged hop prediction for node $v_i$. For convenience, we use $S_{AHP}(V)$ denote the averaged hop predictions for all nodes:
\begin{align}
S_{AHP}(V)=\{S_{AHP}(v_1),S_{AHP}(v_2),...,S_{AHP}(v_i)\}, v_i \in \bm{V}
\label{ahp}
\end{align}

\textbf{Average Hop Prediction+Inferred Variance (HAV)} which integrates the averaged hop predictions and inferred variance. The variance of hop prediction between node $v_i$ and node $v_j$ is defined as followings:
\begin{align}
    \delta (v_i,v_j)= \frac{\sum_t^T (F_{\bm{w}_{t}}(v_i,v_j \lvert \bm{X},\bm{A})-\widetilde{Y}_{hop}(v_i,v_j))^2}{T},
\end{align}
where $v_i,v_j \in \bm{V}$.  For one node, we use the averaged variance of hop predictions between the node and its neighbors to present the anomaly scores of this node.
\begin{align}
    S_{IV}(v_i)=\frac{\sum_{v_j} \delta (v_i,v_j)}{\lvert \mathcal{N}(v_i) \rvert}, v_j \in \mathcal{N}(v_i).
    \label{iv_i}
\end{align}
Similarly, We use $S_{IV}(V)$ to denote the variance for all nodes.
\begin{align}
 S_{IV}(V)=\{S_{IV}(v_1),S_{IV}(v_2),...,S_{IV}(v_i)\}, v_i \in \bm{V}.
 \label{iv}
\end{align}
Finally, we integrate the anomaly scores from the predicted hop counts and the inferred variance together. The expression is defined as follows:
\begin{align}
    S_{HAV}(V)=\frac{S_{AHP}(V)}{\max(S_{AHP}(V))} +  \frac{S_{IV}(V)}{\max(S_{IV}(V))},
    \label{hav}
\end{align}
where $\max(\cdot)$ represents the maximum value of the set. Considering that the predicted hop counts and inferred variance have different scales, we use the maximum to normalize each score. The pseudocode of anomaly scores are summarized in Algorithm~\ref{alg:2} (In Appendix~\ref{algo}).

\section{Experiments}
\label{exp}
In this section, we evaluate our proposed HCM model empirically. We aim to answer the following three research questions:
\begin{itemize}
    \item RQ1: Is the HCM model effective in detecting anomalies on attributed networks?
    \item RQ2: Does SGLD optimization framework benefit the anomaly detection performance of our approach on attributed networks?
    \item RQ3: How do the hyper-parameters in the HCM model affect the anomaly detection performance?
\end{itemize}
We first introduce the datasets and experimental settings. Then we report and analyze the experimental results.

\subsection{Datasets}
We use five real-world attributed networks to evaluate the effectiveness of our proposed method. These networks have been widely used in previous anomaly detection studies~\cite{li2017radar,peng2018anomalous,ding2019deep,gutierrez2019multi,pei2020resgcn}. These networks can be categorized into two types: networks with ground-truth anomaly labels and networks with injected anomaly labels.
\begin{itemize}
    \item For networks with ground-truth anomaly labels, Amazon and Enron\footnote{\url{https://www.ipd.kit.edu/mitarbeiter/muellere/consub/}} have been used. Amazon is a co-purchase network~\cite{muller2013ranking}. The anomalous nodes are defined as nodes having the tag \textit{amazonfail}. Enron is an email network~\cite{metsis2006spam}. Attributes represent content length, number of recipients, etc, and each edge indicates the email transmission between sender and receiver. Spammers are the anomalies.
    \item For networks with injected anomaly labels, we select BlogCatalog, Flickr and ACM\footnote{\url{http://people.tamu.edu/~xhuang/Code.html}}. BlogCatalog is a blog sharing website. A list of tags associated with each user is used as the attributes. Flickr is an image hosting and sharing website. Similarly, tags are the attributes. ACM is a citation network. Words in abstracts are used as attributes. 
\end{itemize}
A brief statistics of these attributed networks are showed in Table~\ref{tb:data} (In Appendix~\ref{DS}). Note that for the data with injected labels, to make a fair comparison, we follow previous studies for anomaly injection~\cite{ding2019deep,pei2020resgcn}. In specific, two anomaly injection methods have been used to inject anomalies by perturbing topological structure and nodal attributes, respectively:
\begin{itemize}
    \item \textbf{Structural anomalies}: we perturb the topological structure of the input network to generate structural anomalies. 
    We randomly select $s$ nodes from the network and then make those nodes fully connected, and then all the $s$ nodes forming the clique are labeled as anomalies. $t$ cliques are generated repeatedly and totally there are $s\times t$ structural anomalies.
    
    \item \textbf{Attribute anomalies}: we perturb the nodal attributes of the input network to generate attribute anomalies. Same to~\cite{ding2019deep,song2007conditional,pei2020resgcn}, we first randomly select $s\times t$ nodes as the attribute perturbation candidates. For each selected node $v_i$, we randomly select another $k$ nodes from the network and calculate the Euclidean distance between $v_i$ and all the $k$ nodes. Then the node with largest distance is selected as $v_j$ and the attributes $\bm{X}_{j,:}$ of node $v_j$ is changed to $\bm{X}_{i,:}$ of node $v_i$. The selected node $v_j$ is regarded as the generated attribute anomaly.
    
\end{itemize}
In the experiments, we set $s=15$ and set $t$ to 10, 15, and 20 for BlogCatalog, Flickr and ACM, respectively which are the same to~\cite{ding2019deep,pei2020resgcn} in order to make a fair comparison. To facilitate the learning process, in our experiments, we use Principal Component Analysis (PCA)~\cite{wold1987principal} to reduce the dimensionality of attributes to 20 following~\cite{ding2019interactive}.

\subsection{Experimental Settings}
In the experiments, we use the HCM model consisting of four convolution layers and two fully connected layers for Amazon and Enron networks, two convolution layers and two fully connected layers for BlogCatalog, Flickr and ACM networks. The units of each convolution layer are 128. The units of the fully connected layer are set to 256 and the classes $C$ respectively. The learning rate  is set to 0.01. The weight decay is set to 5e-8. The default drop ratio $R$, classes $C$ and sampling ratio $S$ are set to 0.2, 4 and 0.3 respectively.

In the experiments, following the previous studies~\cite{li2017radar,peng2018anomalous,ding2019deep,gutierrez2019multi,pei2020resgcn}, we use the area under the receiver operating characteristic curve (ROC-AUC) as the evaluation metric for anomaly detection. We compare the proposed HCM model with the following anomaly detection methods:\textbf{LOF}~\cite{breunig2000lof}, \textbf{AMEN}~\cite{perozzi2016scalable}, \textbf{Radar}~\cite{li2017radar}, \textbf{ANOMALOUS}~\cite{peng2018anomalous},  \textbf{DOMINANT}~\cite{ding2019deep}, \textbf{MADAN}~\cite{gutierrez2019multi} and \textbf{ResGCN}~\cite{pei2020resgcn}.

\begin{table}
\vspace{-0.3in}
\caption{Performance of different anomaly detection methods w.r.t.\ ROC-AUC on Flickr, BlogCatalog, ACM, Amazon and Enron networks.-:No results.}
\label{tb:auc_synthetic}
\centering
\begin{tabular}{l|c|c|c|c|c}
\hline
&Flickr  &BlogCatalog     &ACM &Amazon&Enron                \\\hline
LOF~\cite{breunig2000lof}       & 0.488  & 0.491 &0.473&0.490&0.440 \\ \hline
AMEN~\cite{perozzi2016scalable} & 0.604  & 0.665& 0.533&0.470&0.470 \\ \hline
Radar~\cite{li2017radar}     & 0.728  & 0.725&0.693&0.580&0.650 \\\hline
ANOMALOUS~\cite{peng2018anomalous} & 0.716 &0.728&0.718&0.602&0.695 \\ \hline
DOMINANT~\cite{ding2019deep}  & 0.781 &0.749 &0.749 &0.625&0.685\\ \hline
MADAN~\cite{gutierrez2019multi}& - & - &-&0.680&0.660\\ \hline
ResGCN~\cite{pei2020resgcn}& 0.780 & 0.785 &0.768 &\textbf{0.710}&0.660\\ \hline
HCM-AHP (Ours)  &0.791   &\textbf{0.808} &\textbf{0.806} &0.62 &0.670\\ \hline
HCM-HAV (Ours)  &\textbf{0.792} &0.798       &0.761 &0.708&\textbf{0.715} \\ \hline
\end{tabular}
\vspace{-0.3in}
\end{table}

\subsection{Experimental Results}
We conduct experiments to evaluate the performance of our proposed method and to compare it with several state-of-the-art approaches, to which we refer as baselines, on two different types of networks: networks with ground-truth anomalies and injected anomalies respectively.

\textbf{Results on the networks with known ground-truth anomalies} Results on the network with ground-truth anomalies are showed in Table~\ref{tb:auc_synthetic}. It shows that the HCM-HAV score achieves the best among the baselines on Enron network and comparable results on Amazon. The HCM-AHP score leads to comparable results achieved with DOMINANT and ANOMALOUS, demonstrating the effectiveness of utilizing both global and local information to detect anomalies. The better performance of the HCM-HAV score than the HCM-AHP score indicates that the inferred variance plays a role in finding anomalies in the Amazon and Enron networks. 

\textbf{Results on the networks with injected anomalies} Results on the network with injected anomalies are showed in table~\ref{tb:auc_synthetic}. As can be seen, HCM-HAV and HCM-AHP scores outperform the baseline methods in all cases, which further validates the effectiveness of our method by utilizing both global and local information to detect anomalies. 

\begin{figure}[htb]
\vspace{-0.3in}
\centering
\subfloat[{\small Average $S_{AHP}$}]{\includegraphics[width=0.48\textwidth]{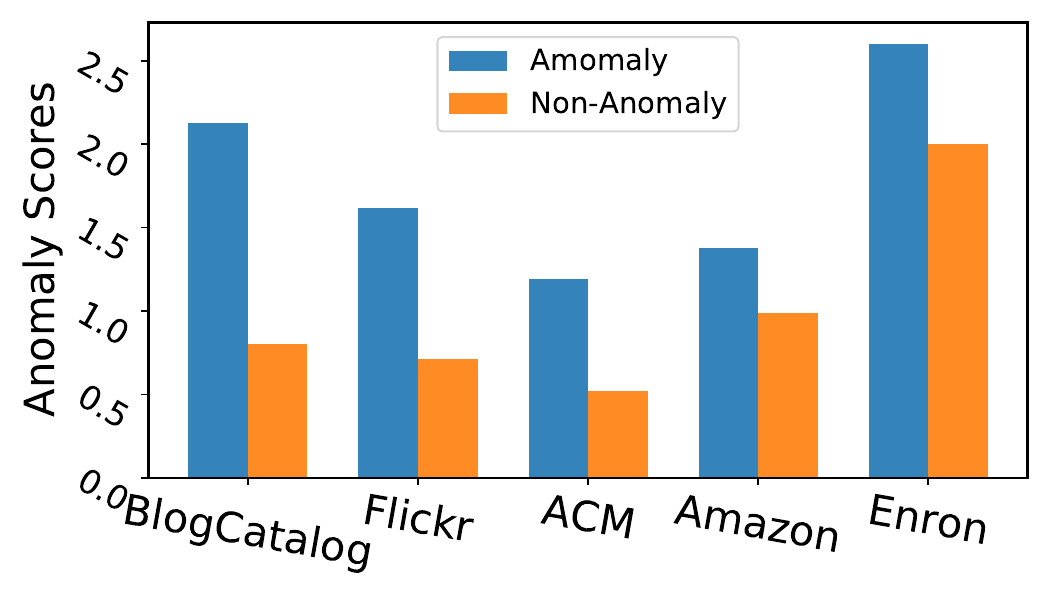}
}
\subfloat[{\small Average $S_{HAV}$}]{\includegraphics[width=0.48\textwidth]{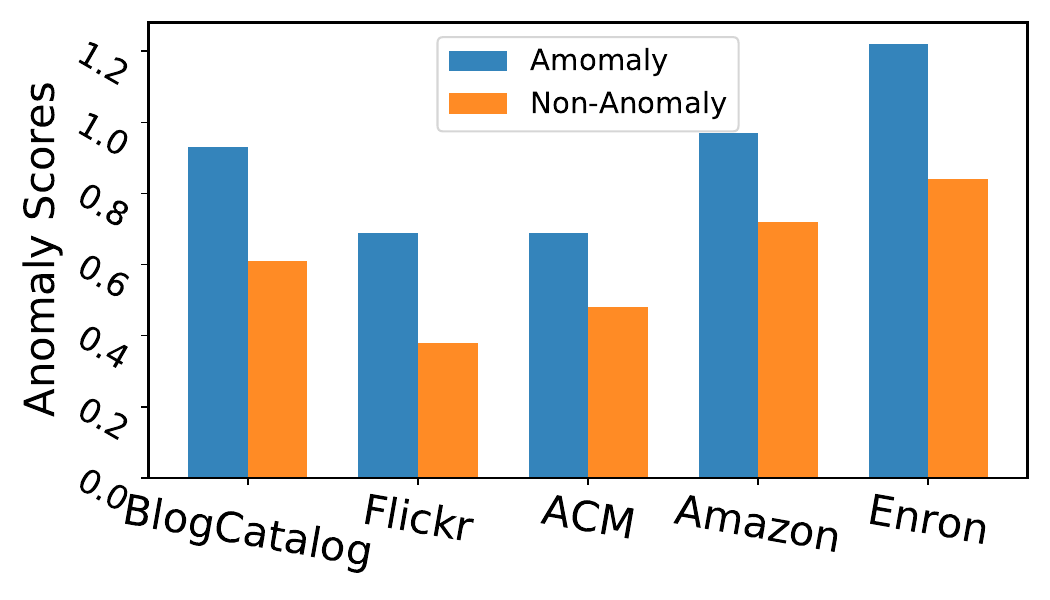}
}
\caption{Average Scores of Anomalous nodes and Non-Anomalous nodes.}
\label{fig:scores}
\vspace{-0.3in}
\end{figure}

The insight of HCM-AHP is that the hop counts between the anomalous nodes and their neighbors are expected to be larger than that between normal nodes and their neighbors. The insight of HCM-HAV score integrates the influence of inferred variance into HCM-AHP. Fig.~\ref{fig:scores} shows that the average HCM-AHP and HCM-HAV among anomalous nodes are bigger than among non-anomalous nodes, which further validates the rationality of the insight. Interestingly, we note that the HCM-HAV score doesn't perform better than HCM-AHP score on BlogCatalog and ACM networks, which indicates that the integration of inferred variance and HCM-AHP score does not constantly improve the performance. However, HCM-HAV can consistently achieve good results on all networks compared to HCM-AHP.

\begin{figure*}[htb]
\centering
\setlength\tabcolsep{1pt}
\settowidth\rotheadsize{Radcliffe Cam}
\begin{tabularx}{0.9\linewidth}{XXX}
   \includegraphics[width=\hsize,valign=m]{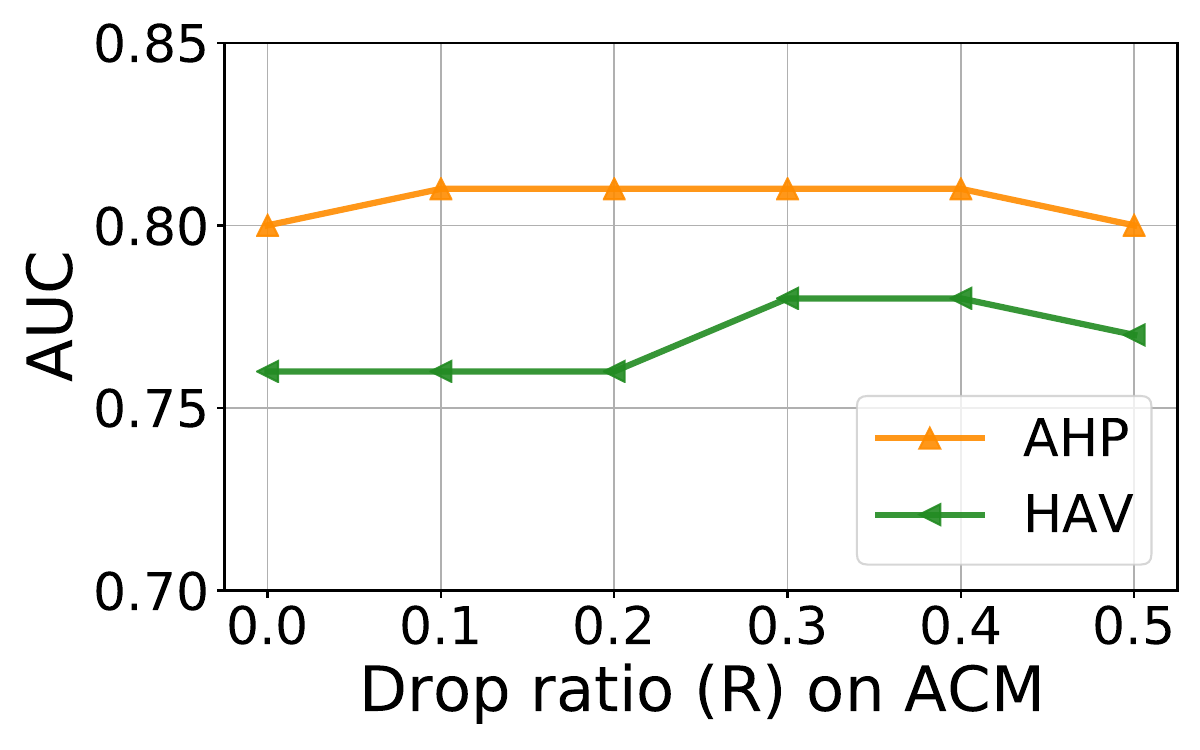}    
                        &   \includegraphics[width=\hsize,valign=m]{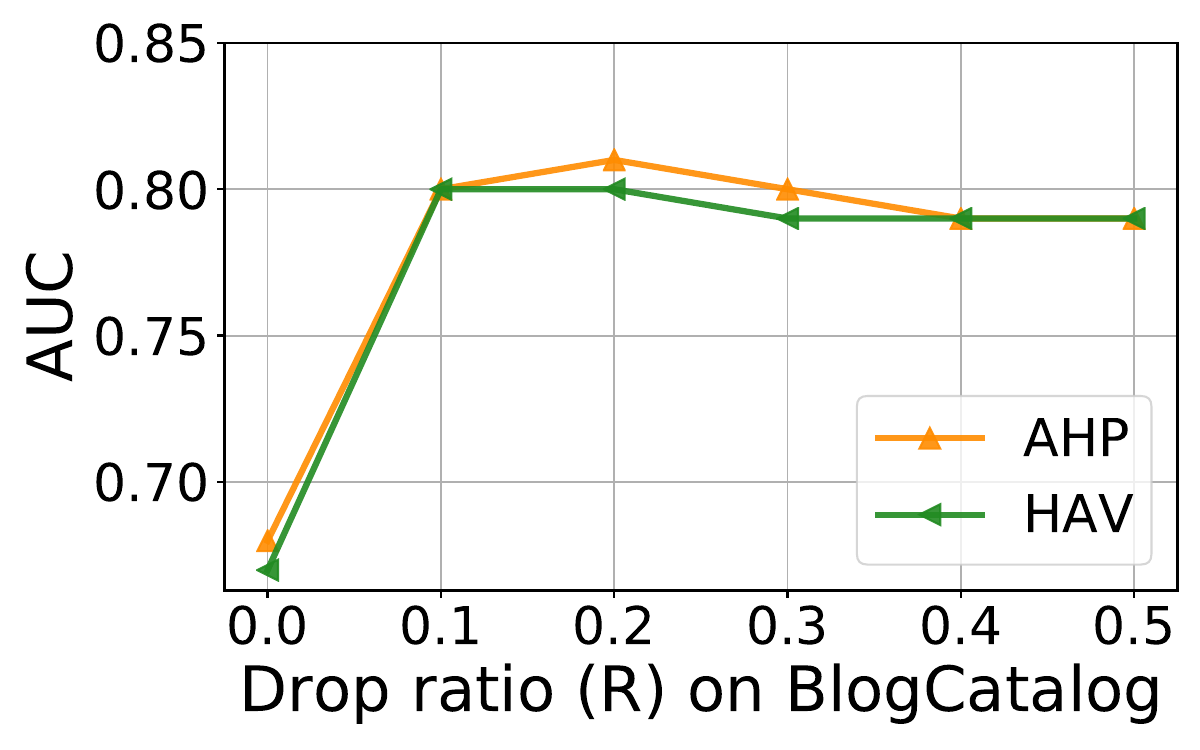}
                        & \includegraphics[width=\hsize,valign=m]{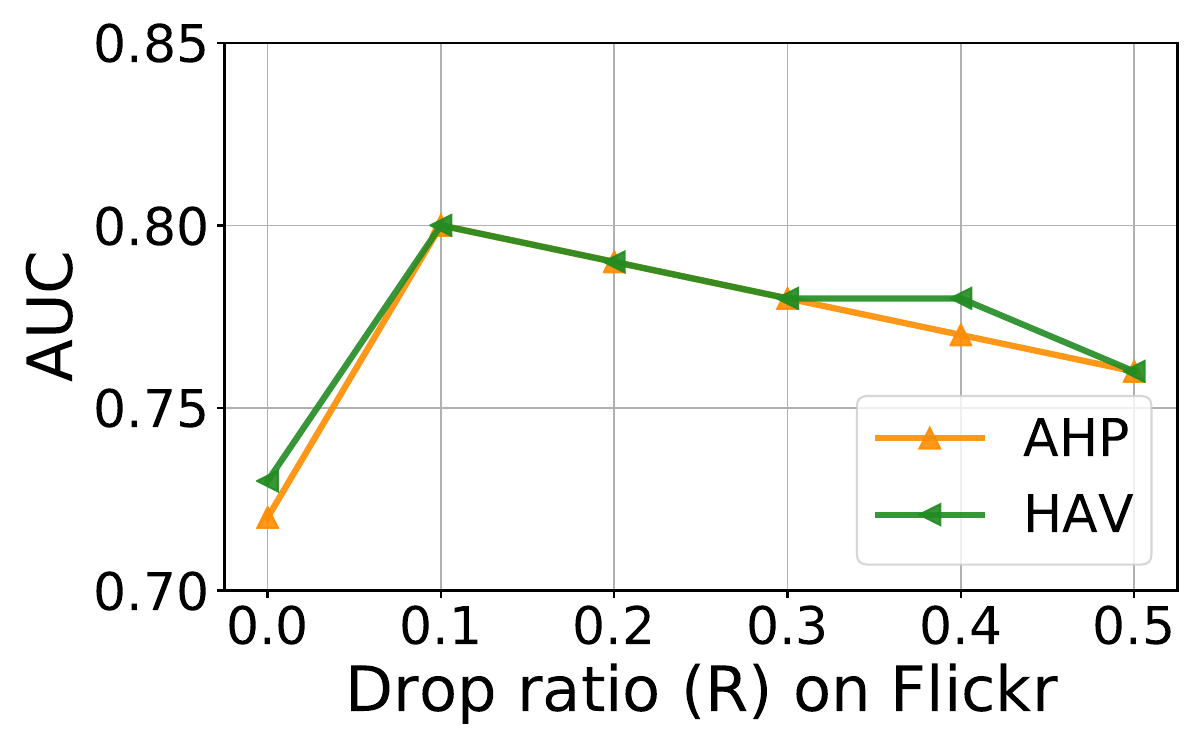}
                        \\[-0em]                        \includegraphics[width=\hsize,valign=m]{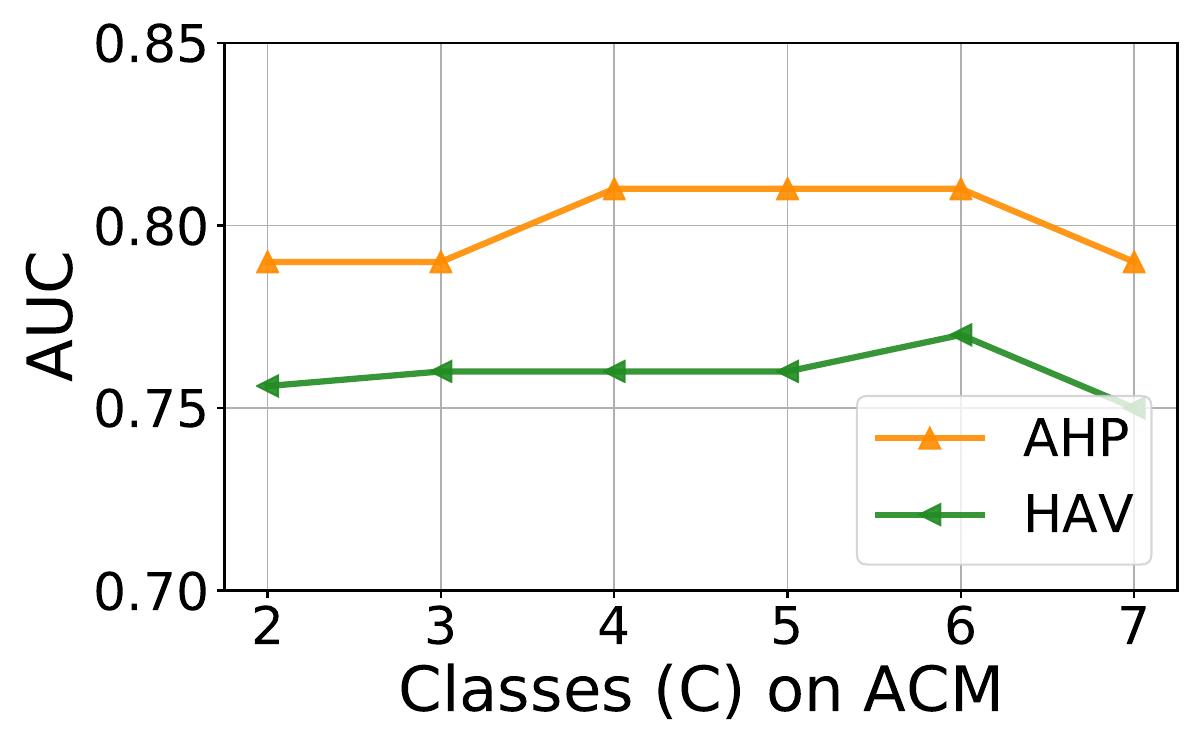}
                        &   \includegraphics[width=\hsize,valign=m]{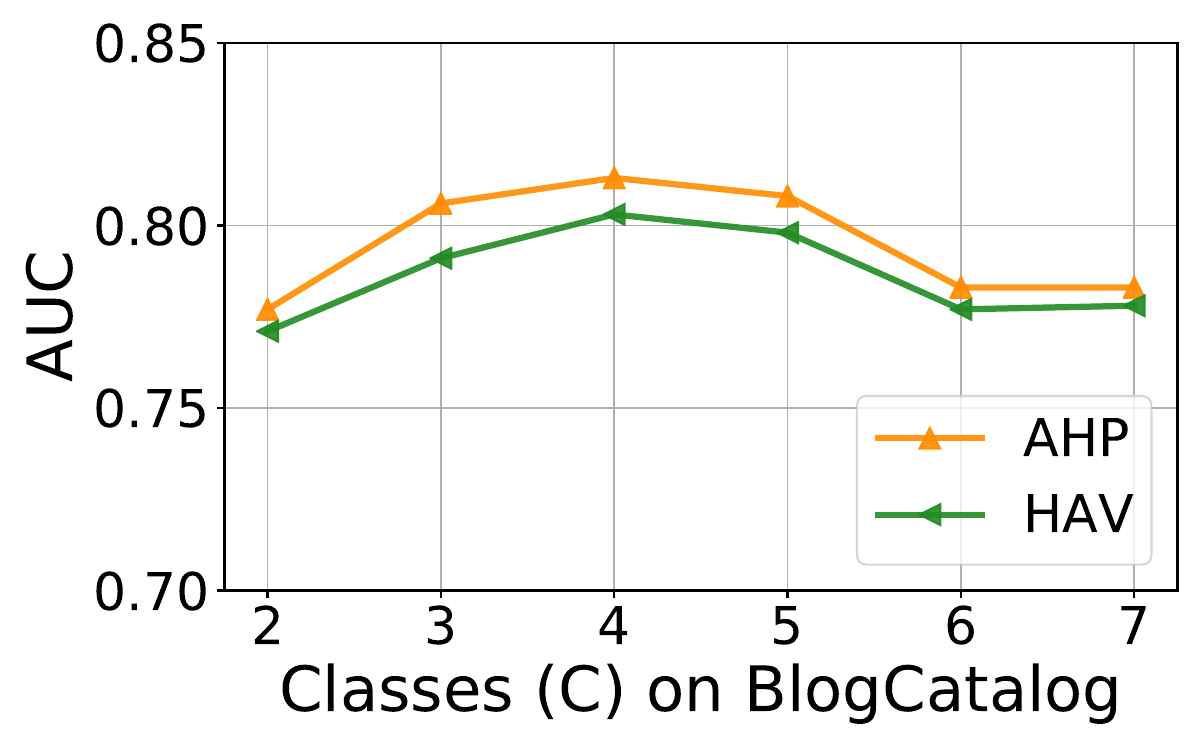}
                        &   \includegraphics[width=\hsize,valign=m]{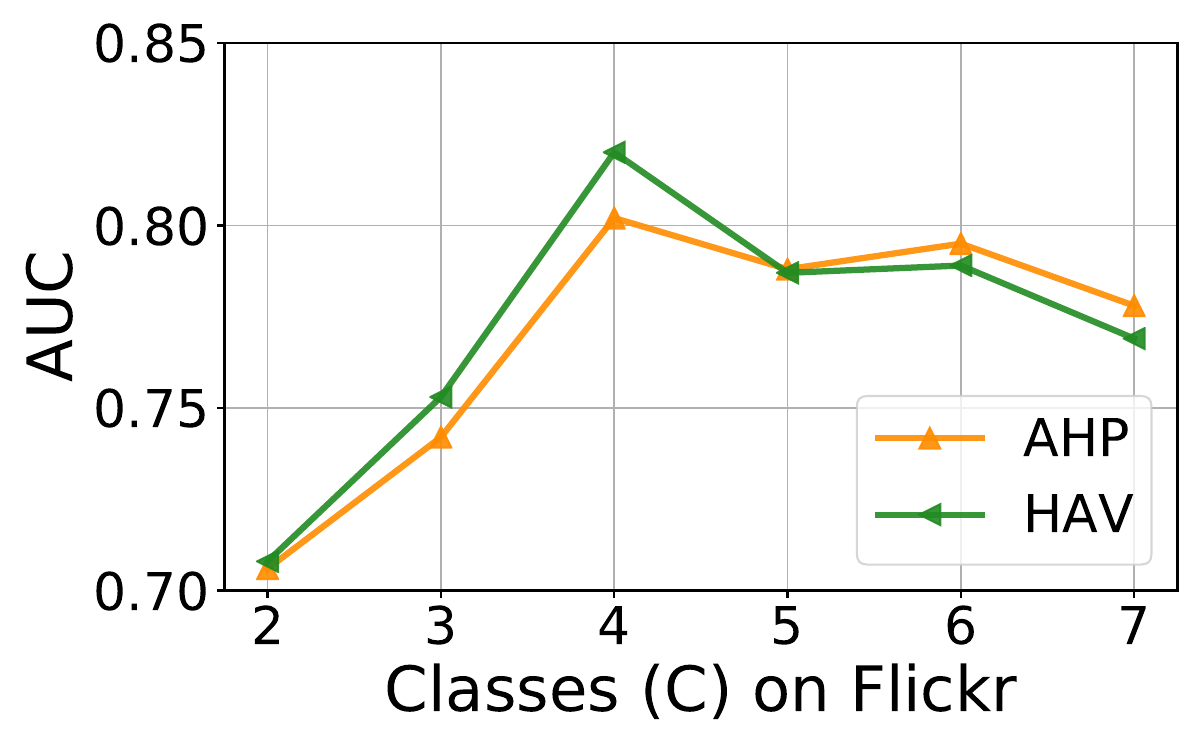}
                        \\[-0em]                        \includegraphics[width=\hsize,valign=m]{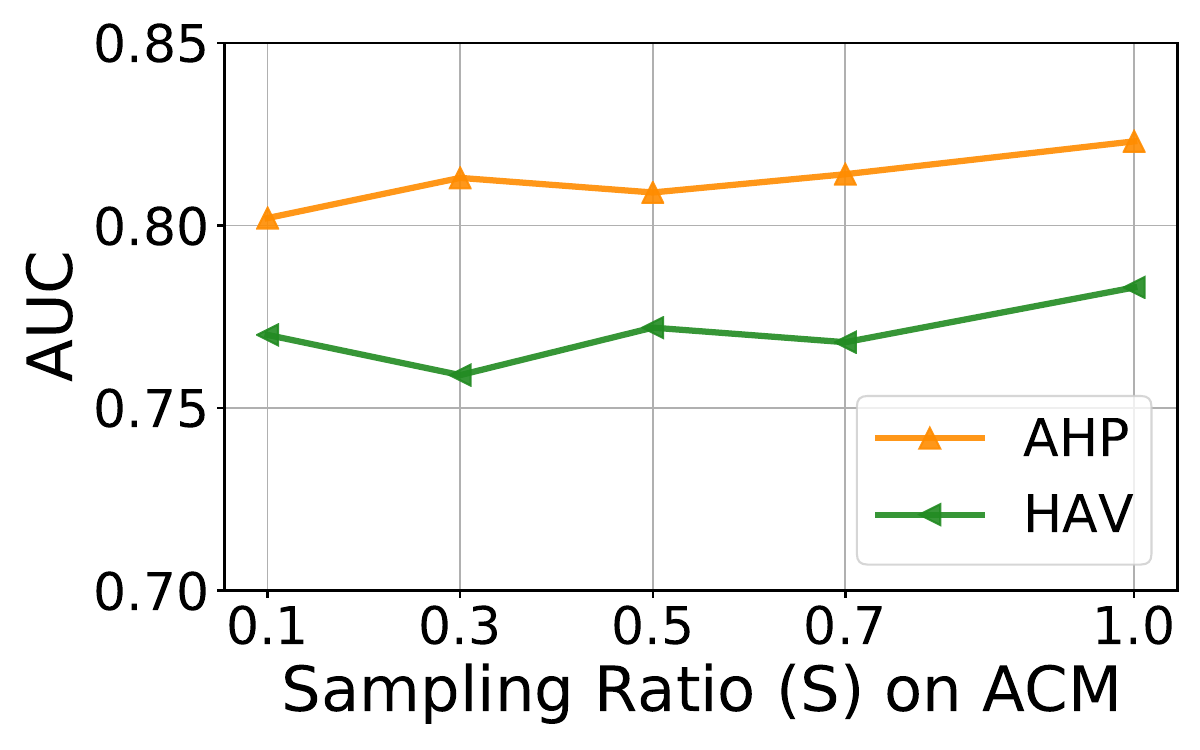}
                        &   \includegraphics[width=\hsize,valign=m]{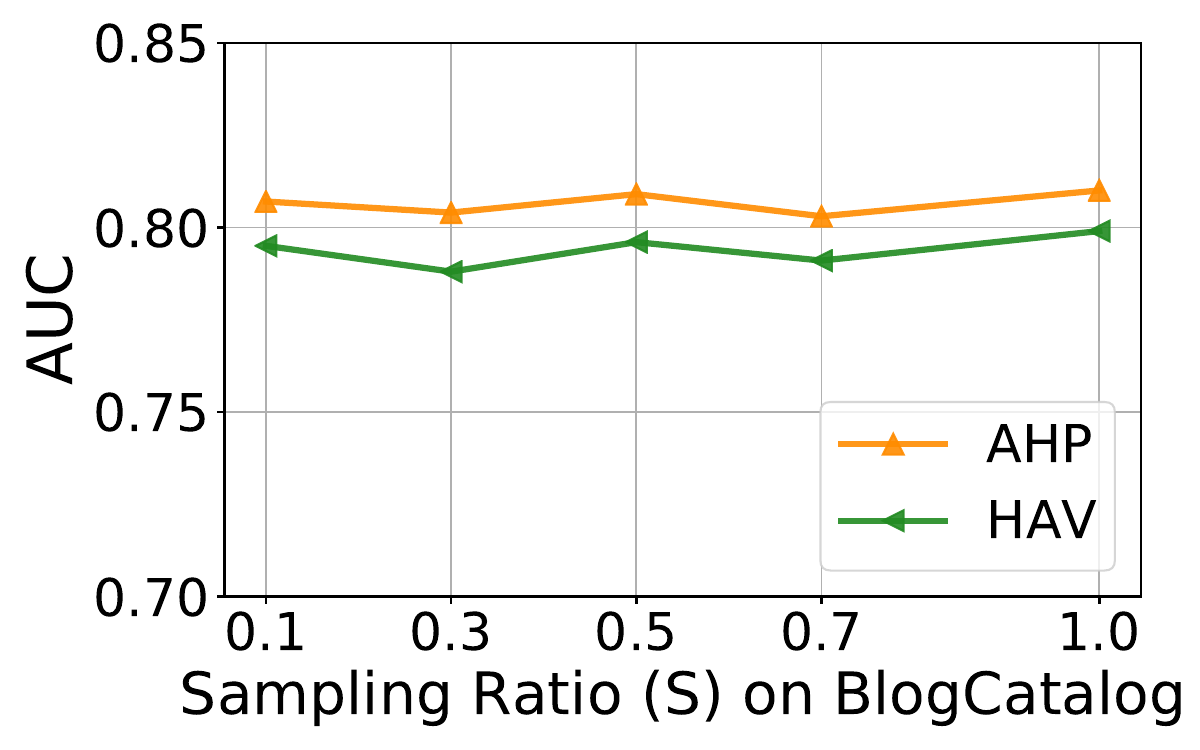}
                        &   \includegraphics[width=\hsize,valign=m]{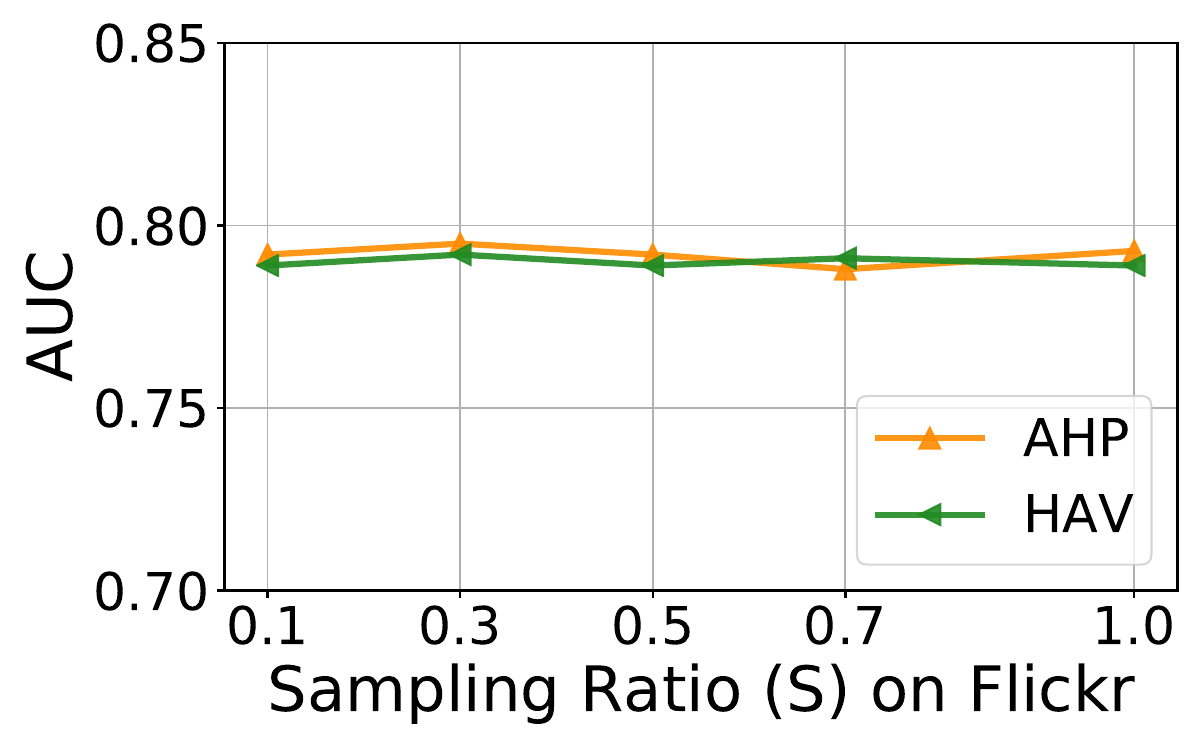}
                        \\[-0em]
\end{tabularx}
    \caption{Influence of the hyper-parameters: Drop ratio ($R$) (ranging from 0 to 0.5), Classes ($C$) (ranging from 2 to 7), Sampling Ratio ($S$) (ranging from 0.1 to 1.0). Experiments are conducted on ACM, BlogCatalog and Flickr respectively. }
\label{fig:parameters}
\vspace{-0.3in}
\end{figure*}

\subsection{Parameter Analysis}
We conducted a series of experiments to study the impact of hyper-parameters on the performance of anomaly detection. There are three hyper-parameters introduced in our proposed method: (1)~drop ratio $R$ in the Preprocess component, which controls the percentages of dropped edges, (2)~classes $C$, which controls the max hop counts that the training set contains, (3)~sampling ratio $S$, that controls the amount of node pairs engaging in the training process of each iteration.

\textbf{Drop Ratio $R$.} We fix classes $C$ and sampling ratio $S$ to 5 and 0.3 respectively. Then we monitor the anomaly performance by ranging drop ratio $R$ from 0 t 0.5 on BlogCatalog, Flickr and ACM datasets. The results are showed in Fig.~\ref{fig:parameters} (the first row). 

From the results, we can see that removing a few edges with low similarity is beneficial in detecting anomalies. Specifically, the experiments show that the performance of anomaly detection increases greatly on BlogCatalog and Flickr datasets when a few  edges are removed. However, it is not encouraged to remove a large amount of edges since the performance of anomaly detection has slightly decreased when a large amount of edges are removed. 

\textbf{Classes $C$.} We set drop ratio $R$ and sampling ratio $S$ to 0.2 and 0.3 respectively. The classes $C$ varies from 2 to 7. Experiments are done on BlogCatalog, Flickr and ACM networks and the results are showed in Fig.~\ref{fig:parameters} (the second row).  

From the figures, we can see that the performance of anomaly detection is improved with the increase of the classes $C$, especially on Flickr and BlogCatalog datasets. The improvement suggests that the global information is beneficial to anomaly detection since the more classes $C$, the more global information is used. 

\textbf{Sampling Ratio $S$.} We fix hyper-parameters $R$ and $C$ to 0.2 and 5 respectively. We set sampling ratio $S$ varying from 0.1 to 1.0. Experiments are done on BlogCatalog, Flickr and ACM networks. The results are showed in Fig.~\ref{fig:parameters} (the third row).

From the results, we can see that the performance of detecting anomalies is insensitive to sampling ratio $S$.

\begin{figure}[htb]
\vspace{-0.3in}
\centering
\subfloat[{\small $S_{AHP}$}]{\includegraphics[width=0.48\textwidth]{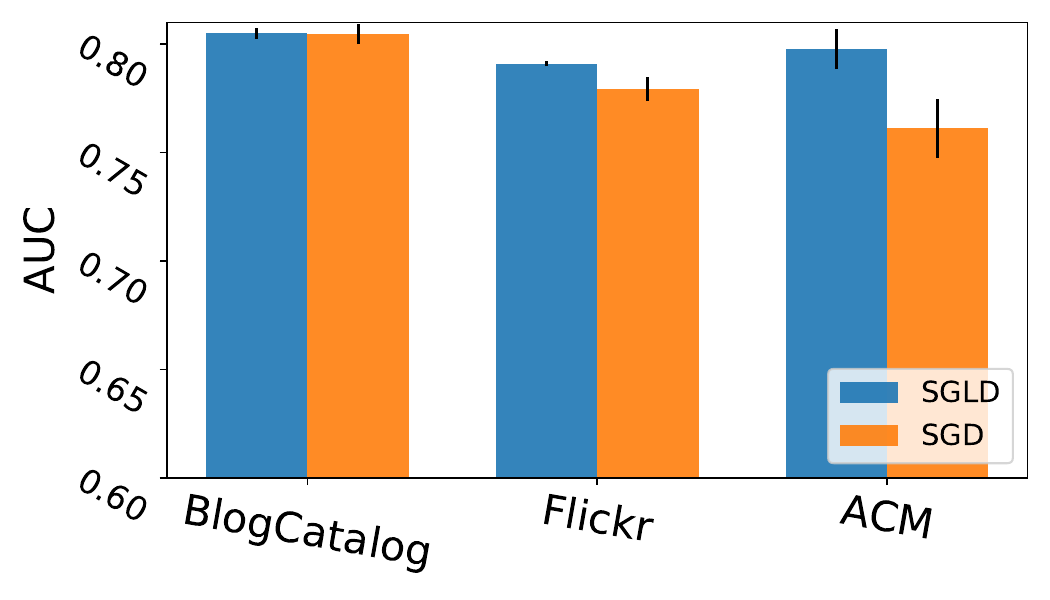}\label{fig:ablation_AHP}
}
\subfloat[{\small $S_{HAV}$}]{\includegraphics[width=0.48\textwidth]{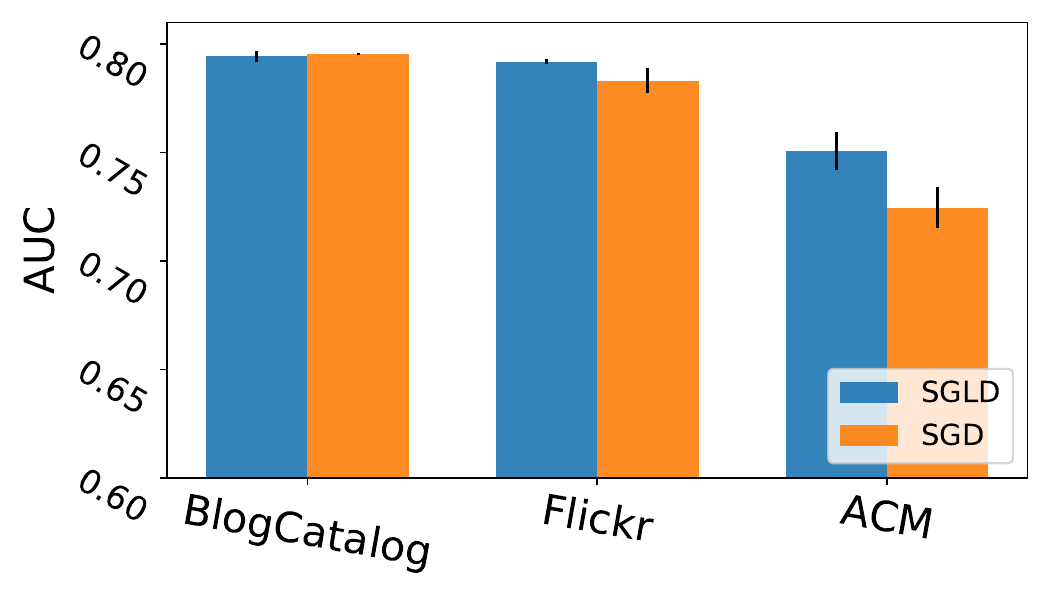}\label{fig:ablation_HAV}
}
\caption{ROC-AUC achieved by $S_{HAV}$ and $S_{AHP}$ under SGLD  and SGD training strategy respectively.}
\label{fig:ablation}
\vspace{-0.1in}
\end{figure}

\begin{wrapfigure}{r}{0.5\linewidth}
	\vspace{-0.5in}
	\begin{center}
		\includegraphics[width=0.5\textwidth]{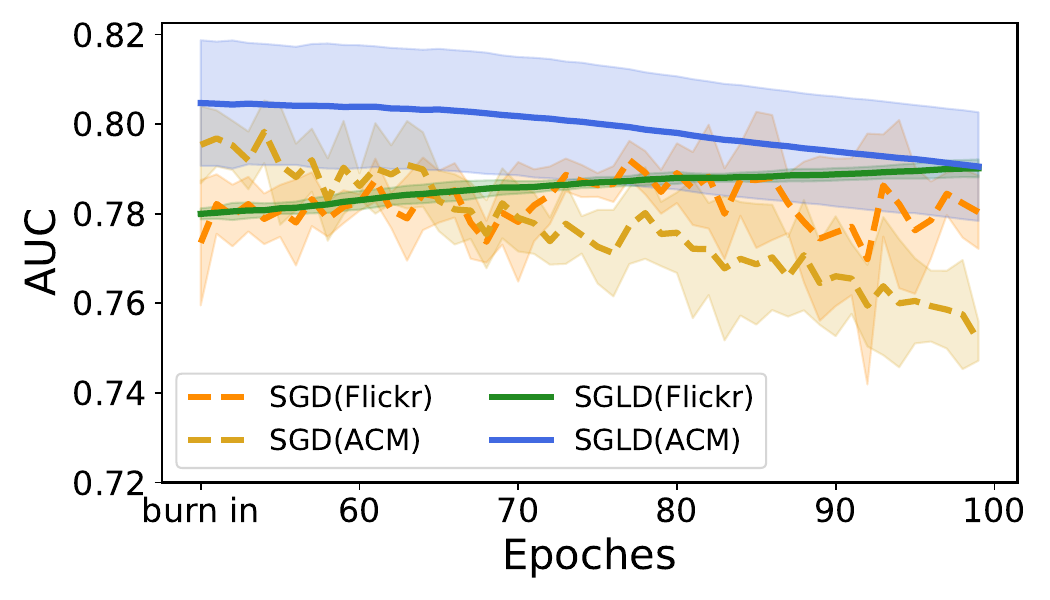}
		\caption{ROC-AUC trends during the training process under SGLD and SGD training strategy respectively. The ROC-AUC value is the mean of 10 times repeated experiments. The filled region along trends is the standard deviation of the 10 times repeated experiments. }
		\label{fig:ablation_training}
	\end{center}
	\vspace{-0.4in}
\end{wrapfigure}
\subsection{Ablation Study}
For the proposed HCM model, SGLD is used instead of stochastic gradient descent (SGD) to optimize the model since we adopt Bayesian learning. Here we conduct experiments to show the benefits of utilizing SGLD to train our proposed model. Specifically, we firstly test the anomaly detection performances of the model optimized by SGD and SGLD respectively. Then we monitor the trend of ROC-AUC during the training process under SGD and SGLD training strategy respectively. The results are showed in Fig.~\ref{fig:ablation_AHP}, Fig.~\ref{fig:ablation_HAV} and Fig.~\ref{fig:ablation_training}. We can observe the following from these figures:
\begin{itemize}
    \item From Fig.~\ref{fig:ablation_AHP} and Fig.~\ref{fig:ablation_HAV}, we see that the ROC-AUC performance achieved by SGLD is better than that by SGD on Flickr and ACM networks, which shows that SGLD based Bayesian learning benefits the anomaly detection on attributed networks.
    \item From Fig.~\ref{fig:ablation_training},  we see that the AUC trend under SGLD is steadier than that under SGD.
\end{itemize}

\section{Conclusion}
\label{conc}
In this paper, we proposed a HCM model based on self-supervised technique for anomaly detection. To the best of our knowledge, it is the first model to take both local and global contextual information into account for anomaly detection. Specifically, we utilize a hop-count based self-supervised task to learn node representation with capturing local and global contextual information. The learned node representation will be used to predict hop counts for arbitrary node pairs as well. Besides, we designed two new anomaly scores for detecting anomalies based on the hop counts prediction via the HCM model. Finally, we introduce SGLD to train the model for capturing uncertainty in learned parameters and avoiding overfitting. The extensive experiments demonstrate: 1)~the consistent effectiveness of the HCM model in anomaly detection on attributed networks that we proposed; 2)~SGLD-based Bayesian learning strategy is beneficial to achieve a better and stable performance in detecting anomalies.


%
%
%
\bibliographystyle{splncs04}
\bibliography{sn-bibliography}

\begin{thebibliography}{10}
\providecommand{\url}[1]{\texttt{#1}}
\providecommand{\urlprefix}{URL }
\providecommand{\doi}[1]{https://doi.org/#1}

\bibitem{breunig2000lof}
Breunig, M.M., Kriegel, H.P., Ng, R.T., Sander, J.: Lof: identifying
  density-based local outliers. In: Proceedings of the 2000 ACM SIGMOD
  international conference on Management of data. pp. 93--104 (2000)

\bibitem{brochier2019link}
Brochier, R., Guille, A., Velcin, J.: Link prediction with mutual attention for
  text-attributed networks. In: Companion Proceedings of The 2019 World Wide
  Web Conference. pp. 283--284 (2019)

\bibitem{ding2019deep}
Ding, K., Li, J., Bhanushali, R., Liu, H.: Deep anomaly detection on attributed
  networks. In: Proceedings of the 2019 SIAM International Conference on Data
  Mining. pp. 594--602. SIAM (2019)

\bibitem{ding2019interactive}
Ding, K., Li, J., Liu, H.: Interactive anomaly detection on attributed
  networks. In: Proceedings of the Twelfth ACM International Conference on Web
  Search and Data Mining. pp. 357--365 (2019)

\bibitem{falih2018community}
Falih, I., Grozavu, N., Kanawati, R., Bennani, Y.: Community detection in
  attributed network. In: Companion Proceedings of the The Web Conference 2018.
  pp. 1299--1306 (2018)

\bibitem{gao2010community}
Gao, J., Liang, F., Fan, W., Wang, C., Sun, Y., Han, J.: On community outliers
  and their efficient detection in information networks. In: Proceedings of the
  16th ACM SIGKDD international conference on Knowledge discovery and data
  mining. pp. 813--822 (2010)

\bibitem{gutierrez2019multi}
Guti{\'e}rrez-G{\'o}mez, L., Bovet, A., Delvenne, J.C.: Multi-scale anomaly
  detection on attributed networks. arXiv preprint arXiv:1912.04144  (2019)

\bibitem{hamilton2017inductive}
Hamilton, W., Ying, Z., Leskovec, J.: Inductive representation learning on
  large graphs. In: Advances in Neural Information Processing Systems. pp.
  1024--1034 (2017)

\bibitem{hendrycks2019using}
Hendrycks, D., Mazeika, M., Kadavath, S., Song, D.: Using self-supervised
  learning can improve model robustness and uncertainty. In: Advances in Neural
  Information Processing Systems. pp. 15663--15674 (2019)

\bibitem{huang2017accelerated}
Huang, X., Li, J., Hu, X.: Accelerated attributed network embedding. In:
  Proceedings of the 2017 SIAM international conference on data mining. pp.
  633--641. SIAM (2017)

\bibitem{jin2020self}
Jin, W., Derr, T., Liu, H., Wang, Y., Wang, S., Liu, Z., Tang, J.:
  Self-supervised learning on graphs: Deep insights and new direction. arXiv
  preprint arXiv:2006.10141  (2020)

\bibitem{johnson1977efficient}
Johnson, D.B.: Efficient algorithms for shortest paths in sparse networks.
  Journal of the ACM (JACM)  \textbf{24}(1),  1--13 (1977)

\bibitem{kendall2017uncertainties}
Kendall, A., Gal, Y.: What uncertainties do we need in bayesian deep learning
  for computer vision? In: Advances in neural information processing systems.
  pp. 5574--5584 (2017)

\bibitem{kipf2016semi}
Kipf, T.N., Welling, M.: Semi-supervised classification with graph
  convolutional networks. arXiv preprint arXiv:1609.02907  (2016)

\bibitem{li2018streaming}
Li, J., Cheng, K., Wu, L., Liu, H.: Streaming link prediction on dynamic
  attributed networks. In: Proceedings of the Eleventh ACM International
  Conference on Web Search and Data Mining. pp. 369--377 (2018)

\bibitem{li2017radar}
Li, J., Dani, H., Hu, X., Liu, H.: Radar: Residual analysis for anomaly
  detection in attributed networks. In: IJCAI. pp. 2152--2158 (2017)

\bibitem{liang2018semi}
Liang, J., Jacobs, P., Sun, J., Parthasarathy, S.: Semi-supervised embedding in
  attributed networks with outliers. In: Proceedings of the 2018 SIAM
  International Conference on Data Mining. pp. 153--161. SIAM (2018)

\bibitem{meng2019co}
Meng, Z., Liang, S., Bao, H., Zhang, X.: Co-embedding attributed networks. In:
  Proceedings of the Twelfth ACM International Conference on Web Search and
  Data Mining. pp. 393--401 (2019)

\bibitem{metsis2006spam}
Metsis, V., Androutsopoulos, I., Paliouras, G.: Spam filtering with naive
  bayes-which naive bayes? In: CEAS. vol.~17, pp. 28--69. Mountain View, CA
  (2006)

\bibitem{muller2013ranking}
M{\"u}ller, E., S{\'a}nchez, P.I., M{\"u}lle, Y., B{\"o}hm, K.: Ranking outlier
  nodes in subspaces of attributed graphs. In: 2013 IEEE 29th International
  Conference on Data Engineering Workshops (ICDEW). pp. 216--222. IEEE (2013)

\bibitem{pei2015nonnegative}
Pei, Y., Chakraborty, N., Sycara, K.: Nonnegative matrix tri-factorization with
  graph regularization for community detection in social networks. In:
  Twenty-Fourth International Joint Conference on Artificial Intelligence
  (2015)

\bibitem{pei2020resgcn}
Pei, Y., Huang, T., van Ipenburg, W., Pechenizkiy, M.: Resgcn: Attention-based
  deep residual modeling for anomaly detection on attributed networks. arXiv
  preprint arXiv:2009.14738  (2020)

\bibitem{peng2020self}
Peng, Z., Dong, Y., Luo, M., Wu, X.M., Zheng, Q.: Self-supervised graph
  representation learning via global context prediction. arXiv preprint
  arXiv:2003.01604  (2020)

\bibitem{peng2018anomalous}
Peng, Z., Luo, M., Li, J., Liu, H., Zheng, Q.: Anomalous: A joint modeling
  approach for anomaly detection on attributed networks. In: IJCAI. pp.
  3513--3519 (2018)

\bibitem{perozzi2016scalable}
Perozzi, B., Akoglu, L.: Scalable anomaly ranking of attributed neighborhoods.
  In: Proceedings of the 2016 SIAM International Conference on Data Mining. pp.
  207--215. SIAM (2016)

\bibitem{perozzi2014focused}
Perozzi, B., Akoglu, L., Iglesias~S{\'a}nchez, P., M{\"u}ller, E.: Focused
  clustering and outlier detection in large attributed graphs. In: Proceedings
  of the 20th ACM SIGKDD international conference on Knowledge discovery and
  data mining. pp. 1346--1355 (2014)

\bibitem{song2007conditional}
Song, X., Wu, M., Jermaine, C., Ranka, S.: Conditional anomaly detection. IEEE
  Transactions on knowledge and Data Engineering  \textbf{19}(5),  631--645
  (2007)

\bibitem{welling2011bayesian}
Welling, M., Teh, Y.W.: Bayesian learning via stochastic gradient langevin
  dynamics. In: Proceedings of the 28th international conference on machine
  learning (ICML-11). pp. 681--688 (2011)

\bibitem{wold1987principal}
Wold, S., Esbensen, K., Geladi, P.: Principal component analysis. Chemometrics
  and intelligent laboratory systems  \textbf{2}(1-3),  37--52 (1987)

\bibitem{wu2019adversarial}
Wu, H., Wang, C., Tyshetskiy, Y., Docherty, A., Lu, K., Zhu, L.: Adversarial
  examples for graph data: deep insights into attack and defense. In:
  Proceedings of the 28th International Joint Conference on Artificial
  Intelligence. pp. 4816--4823. AAAI Press (2019)

\bibitem{you2020does}
You, Y., Chen, T., Wang, Z., Shen, Y.: When does self-supervision help graph
  convolutional networks? arXiv preprint arXiv:2006.09136  (2020)

\end{thebibliography}


\begin{thebibliography}{8}
\bibitem{ref_article1}
Author, F.: Article title. Journal \textbf{2}(5), 99--110 (2016)

\bibitem{ref_lncs1}
Author, F., Author, S.: Title of a proceedings paper. In: Editor,
F., Editor, S. (eds.) CONFERENCE 2016, LNCS, vol. 9999, pp. 1--13.
Springer, Heidelberg (2016). \doi{10.10007/1234567890}

\bibitem{ref_book1}
Author, F., Author, S., Author, T.: Book title. 2nd edn. Publisher,
Location (1999)

\bibitem{ref_proc1}
Author, A.-B.: Contribution title. In: 9th International Proceedings
on Proceedings, pp. 1--2. Publisher, Location (2010)

\bibitem{ref_url1}
LNCS Homepage, \url{http://www.springer.com/lncs}. Last accessed 4
Oct 2017
\end{thebibliography}
%

\appendix

\onecolumn
\section{Pseudocode}\label{algo}

\begin{algorithm}
\caption{HCM model Training}
\label{alg:1}
\begin{algorithmic}[1]
\Require
Initialize a attribute network $\mathcal{G}$ with nodal attributes $\bm{X}$ and network topology relations $\bm{A}$, with hyper-parameters drop ratio $R$, classes $C$, sampling ratio $S$, step size $\epsilon$; HCM model $F_{\bm{w}}(\bm{X},\bm{A})$; the prior distribution of the parameters p($\bm{w}$).
\Ensure
the trained HCM model $F_{\bm{w}}(\bm{X},\bm{A})$
\State Drop edges using Eq.~\ref{jar} or Eq.~\ref{Sim}
\State Generate true hop counts $Y_{hop}$ based on adjacency matrix with dropped edges using Dijstra search algorithm. 
\State Initialize the parameters of HCM model $\bm{w}$

\For{each epoch i}
    \State $\hat{Y}_{hop}=F_{\bm{w}_{i}}(\bm{X},\bm{A})$
    \State Get $\eta_i$ $\sim$ $\mathcal{N}(0,\epsilon)$
    \State $\delta_{\bm{w}_{i}}=\frac{\epsilon}{2} (\nabla_{\bm{w}_{i}} \mathcal{L}_{self}(\bm{w}_i,\bm{A},\bm{X})+\nabla_{\bm{w}_{i}}log p(\bm{w}_{i}))+\eta_i$
    \State $\bm{w}_{i+1}= \bm{w}_{i}-\delta_{\bm{w}_{i}}$
\EndFor
\\
\Return $F_{\bm{w}}(\bm{X},\bm{A})$
\end{algorithmic}
\end{algorithm}

\begin{algorithm}
\caption{HCM model Inference for Anomaly Scores}
\label{alg:2}
\begin{algorithmic}[1]
\Require
Initialize a attribute network $\mathcal{G}$ with nodal attributes $\bm{X}$ and network topology relations $\bm{A}$; the HCM model $F_{\bm{w}}(\bm{X},\bm{A})$; the prior distribution of the parameters p($\bm{w}$).
\Ensure
AHP score $S_{AHP}(V)$; HAV score $S_{HAV}(V)$
\State $Y \leftarrow \{\}$
\For{$t=0$ to $T$}
    \State $\hat{Y}_{hop}=F_{\bm{w}_{t}}(\bm{X},\bm{A})$
    \State $Y \leftarrow Y+\hat{Y}_{hop}$
    \State Get $\eta_t$ $\sim$ $\mathcal{N}(0,\epsilon)$
    \State $\delta_{\bm{w}_{t}}=\frac{\epsilon}{2} (\nabla_{\bm{w}_t} \mathcal{L}_{self}(\bm{w}_t,\bm{A},\bm{X})+\nabla_{\bm{w}_{t}}log p(\bm{w}_{t}))+\eta_t$
    \State $\bm{w}_{t+1}= \bm{w}_t-\delta_{\bm{w}_{t}}$
\EndFor

\State Get $\widetilde{Y}_{hop}$ by averaging $Y$
\State Get $\delta$ by calculating the variance of $Y$
\State Get $S_{AHP}(V)$ using Eq.~\ref{ahp_i} and Eq.~\ref{ahp}
\State Get $S_{IV}(V)$ using Eq.~\ref{iv_i} and Eq.~\ref{iv}
\State Get $S_{HAV}(V)$ using Eq.~\ref{hav}\\
\Return $S_{AHP}(V)$ and $S_{HAV}(V)$
\end{algorithmic}
\end{algorithm}

\section{Dataset Statistics}\label{DS}

\begin{table*}
\centering
\caption{Statistics of networks.}
\label{tb:data}
\begin{tabular}{l|c|c|c|c|c}
\hline
           & \multicolumn{2}{c|}{ground-truth anomaly} & \multicolumn{3}{c}{injected anomaly} \\ \hline
Dataset   & Amazon              & Enron               & BlogCatalog    & Flickr    & ACM      \\ \hline
\# nodes      & 1,418               & 13,533              & 5,196          & 7,575     & 16,484   \\ \hline
\# edges      & 3,695               & 176,987             & 171,743        & 239,738   & 71,980   \\ \hline
\# attributes & 28                  & 20                  & 8,189          & 12,074    & 8,337    \\ \hline
\# anomalies  & 28                  & 5                   & 300            & 450       & 600      \\ \hline
\end{tabular}
\end{table*}

\section{Evaluation Metrics}
In the experiments, we use the area under the receiver operating characteristic curve (ROC-AUC) as the evaluation metric for anomaly detection as it has been widely used in previous studies~\cite{li2017radar,peng2018anomalous,ding2019deep,gutierrez2019multi,pei2020resgcn}. ROC-AUC can quantify the trade-off between true positive rate (TP) and false positive rate (FP) across different thresholds. The TP is defined as the detection rate, i.e. the rate of true anomalous nodes correctly identified as anomalous, whereas the FP is the false alarm rate, i.e. rate of normal nodes identified as anomalous~\cite{gutierrez2019multi}.

\section{Complexity Analysis}
Our model is composed with tree components. The first component is Preprocess with two operations:Drop Edges and Generating Labels. The complexity of Drop edges is  $O(\lvert E \rvert)$ since it will iterate over all edges. Generating Labels is based on Dijkstra algorithm, so its complexity is $O((\lvert E\rvert+\lvert V \rvert)log{\lvert V \rvert})$~\cite{johnson1977efficient}. The second component is a GCN and its complexity is $O(\lvert E\rvert \cdot d \cdot \lvert F \rvert)$~\cite{pei2020resgcn}, where $d$ is the dimensions of attributes and $\lvert F\rvert$ is the summation of feature maps among all GCN layers. The third component is MLP, which is composed of fully connected layers. For simplicity, suppose that there are $s$ training samples, $t$ input features, $k$ hidden layers and each has $h$ neurons, and $C$ output classes. Its complexity is $O(s \cdot t \cdot h^k \cdot C)$. Therefore, the overall complexity is $O(\lvert E \rvert+(\lvert E\rvert+\lvert V\rvert)log{\lvert V \rvert}+(\lvert E\rvert \cdot d \cdot \lvert F \rvert+s \cdot t \cdot h^k \cdot C)\cdot i)$ where $i$ is the number of epochs.   

By comparison, we also show the complexity of all baselines that used in this paper in Table~\ref{tb:complexity}.

\begin{table*}
\centering
\caption{Comparison of different anomaly detection methods in complexity.}
\label{tb:complexity}
\begin{tabular}{l|c}
\hline
Method &Complexity\\
\hline
LOF & $o(|V|log|V|)$\\
AMEN &$o(|c|^2 d+|E|d)$\\
Radar &$o(id|V|^2+i|V|^3)$\\
ANOMALOUS &$o(id|V|^2)$\\
Dominant &$o(|E|d|F|+|V|^2)$\\
MADAN & $o(|V|^2)$\\
ResGCN &$o(kd|F|+|V|^2)$\\
\hline
\end{tabular}
\vspace{-0.2in}
\end{table*}

\end{document}